\documentclass[preprint,12pt]{elsarticle}

\usepackage{amssymb}
\usepackage{amsmath}
\usepackage{ulem}
\usepackage[greek,english]{babel}

\journal{International Journal of Human-Computer Studies}

\begin{document}
\emergencystretch 1em

\begin{frontmatter}



\title{The Virtual Reality Koinos Method: Analysis of Symmetrical Dyadic Collaboration in Virtual Reality from the perspective of communication models}


\author[label1]{Eloïse Minder\corref{cor1}} 
\ead{eloise_jeanne_claude.minder@ensam.eu}
\cortext[cor1]{Corresponding author.}

\author[label2]{Sylvain Fleury} 
\author[label2,label3]{Solène Neyret} 
\author[label1]{Jean-Rémy Chardonnet} 

\affiliation[label1]{organization={Arts et Metiers Institute of Technology, LISPEN},
            addressline={11 rue Georges Maugey}, 
            city={Chalon-sur-Saône},
            country={France}}

\affiliation[label2]{organization={Arts et Metiers Institute of Technology, LAMPA},
            addressline={rue Marie Curie}, 
            city={Changé},
            country={France}}

\affiliation[label3]{organization={iCube, Strasbourg University},
            addressline={300 boulevard Sébastien Brant}, 
            city={Illkirch},
            country={France}}
            
\begin{abstract}
Understanding which factors influence co-presence in Virtual Reality can help develop more qualitative social interactions, or social interactions that experienced similar sensations, emotions, and feelings to those generated during Face-to-Face interactions. Co-presence has been studied since the beginning of Virtual Reality (VR); however, no consensus has been established on the factors that influence it, except the consensus on the definition of "being there together" within the Virtual Environment. In this paper, we introduce the Koinos method to explain social interactions in VR through communication models, (i) theoretically, and (ii) on two VR symmetrical dyadic collaborative experiments that change the virtual partner's social and physical representations. These analyses lead us to propose an equation that predict and helps manage the sense of co-presence in VR.
\end{abstract}


\begin{highlights}
\item The use of communication models helps to better understand social interaction dynamics in Virtual Reality (VR).
\item This work proposes the Koinos method: a method adapting communication models with the specificities of VR, with the aim of analyzing the dynamics of social interactions in VR.
\item The analysis of two symmetrical dyadic collaborative VR experiments with the Koinos method indicates that avatar appearance does not affect the co-presence level and task completion. 
\item Social representations associated with the virtual collaborator guide the behavior and the perceived quality of social interactions, resulting in significantly lower co-presence with an \textit{AI}-controlled avatar than with a \textit{Human}-controlled avatar.
\end{highlights}


\end{frontmatter}



\section{Introduction}
Among the Computer-Mediated Communication (CMC) devices, Virtual Reality (VR) can be seen as a medium that allows for reproducing social interactions qualia, compared to the Face-to-Face (FtF) interactions. A quale (singular of qualia) is a set of emotions, sensations, and feelings, created and updated from different experiences. Therefore, the qualia of the FtF interactions are continuously updating, and social interactions are central to our daily life. Since our first days, according to Meltzoff \& Moore \citep{meltzoff1977imitation}, the interactions with others allow us to construct ourselves. In VR, to evaluate how much social interactions are similar in VR and FtF, a common measure used is the co-presence level. Defined as the feeling of “being there together” in the Virtual Environment (VE) (Gallagher \& Frith \citep{gallagher2003functional}), understanding which factors could influence it is a primary subject to allow more qualitative social interactions in VR. However, the literature has not provided precise results on this subject, due to the complexity and the cross-disciplinary fields involved.

The study of communication models, such as the social interaction dynamics with Newcomb’s ABX model, Newcomb \citep{newcomb1953approach}, and the Cooperative Model of Human Communication of Tomasello \citep{tom2008}, could help better understand what could influence co-presence in VR. 

In this theoretical study, we introduce the Koinos method by first applying these communication models to VR situations, and then analyzing two VR experiments that use the same collaborative task but with different partner entities and appearances, using the communication models. These analyses lead us to propose a VR co-presence equation that could help predict and manage better the feeling of co-presence, and therefore could help create more qualitative social interaction VR applications.

\section{Related work}
\begin{figure}
    \centering
    \includegraphics[width=0.7\linewidth]{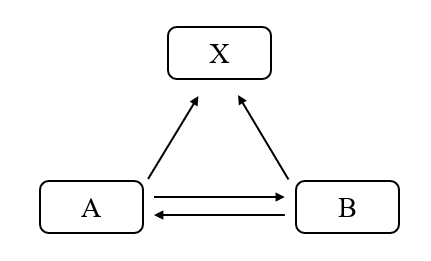}
    \caption{Newcomb’s ABX Model \citep{newcomb1953approach}. A and B can represent individuals and X represents a message transmitted into a context. The double arrow between A and B represents the reciprocal influence, the verbal and non-verbal communication as well as the relationship between them.}
    \label{fig:newcomb}
\end{figure}

Social interactions involve interpersonal communication, which occurs within a social context. Indeed, a conversation will not be conducted in the same way depending on the place it happens, for instance, or the subject of a meeting. In addition, to transmit a message, both verbal and non-verbal communications enable smooth social interaction, allowing us to understand each other, and to take into account the internal states of others. The ABX model of Newcomb \citep{newcomb1953approach} represents these complex social interactions by representing both the relationships between the individuals and the social context (Figure \ref{fig:newcomb}).

A and B can represent two individuals; the double arrow represents the reciprocal influence between the two, encompassing bot verbal and non-verbal communication, as well as the relationship between them. X can represent the topic discussed, such as a third person, or any subject. This model shows that A and B have a "simultaneous orientation" to each other and to X. The theory of Newcomb \citep{newcomb1953approach} indicates that a certain balance exists in this triangle; and that a change in one element changes the entire social interaction dynamics, leading to the emergence of a new equilibrium. For example, we can see A and B as coworkers, exchanging about a work subject. If A receives a promotion and becomes the chief of B, the relationship between A and B will change; therefore, the conversation about the same work subject will also change. 

The Cooperative Model of Human Communication by Tomasello \citep{tom2008} offers insights and complementary information about the interpersonal interaction. This theory considers the individual goals and values in the communication act. To achieve these goals, individuals may need the help, exchange of information, or sharing of attitudes with others, which is the social intention. The social intention is then communicated through the communication intention, allowing for the sharing of common goals. Therefore, the use of references from the external world can be used to communicate the message and the intention, which is the referential intention. Based on norms of cooperation, which indicate how to cooperate with others, the other will try to understand why communication is attempted, and the cooperative reasoning will help understand that. Therefore, to understand the referential intention element, an individual will base his/her reasoning on the \textbf{common ground}, the elements both individuals share. Finally, the individual will decide whether to cooperate or not.

Tomasello proposed that all human beings use this process in communication. Therefore, one factor that influences the understanding in communication is the common ground. The common ground corresponds to elements that interactors have in common. Tomasello \citep{tom2008} describes three levels to evaluate if the common ground is strong or not: 
\begin{itemize}
    \item The common ground is based on the immediate perceptual environment or on a shared experience from the past; 
    \item The situation in which the common ground takes place is based on a top-down process (from brain to action, e.g., focus on the same goal) or a bottom-up process (external signals come to the brain, e.g., hear a noise together);
    \item And the common ground is based on explicit common cultural knowledge (norms, laws, customs) or based on implicit common knowledge (knowing a person that both know the other knows). 
\end{itemize}

Tomasello \citep{tom2008} explains that the common ground is firmer when based on a shared previous experience, in the case of a top-down process, and when it is based on implicit knowledge. Therefore, the “strength” of the common ground depends on the relationship between interactors, on the situation, and on the common knowledge.

In addition, Tomasello \citep{tom2008} indicates that the more the persons share a strong common ground, the less the use of communication elements will be necessary to understand and communicate with one another. We can cite the case of a wife and a husband, twins, or best friends, for instance, who can understand each other at just a glance. Therefore, to understand the communication intention of someone with a very low common ground, more communication will be necessary, both verbal and non-verbal (e.g., asking for a glass of water in a plane to a server from another country).

Previous works have used the common ground perspective to bring new insights to the understanding of social interaction dynamics in social VE. Axelsson et al. \citep{axelsson2003communication} compared symmetric and asymmetric hardware setups (i.e., a CAVE system and computers) on the communication between participants. By comparing the results with the different setups, the authors indicate four factors that allow for better understanding of the others in a collaborative task: hearing the partner, seeing the partner and the partner’s interacted objects, sharing the view of the environment with the partner, and understanding how the partner interacts with the VE. This study supports Tomasello’s theory \citep{tom2008}, which indicates that when individuals do not know each other (i.e., low common ground), more complex communication features are necessary to understand the intentions of others.

It is possible to establish common ground with someone who is unknown in a VE, provided an affordance exists between the task objectives and the interactions within the VE, according to Shami et al. \citep{shami2011places}.

In addition, the model proposed by Roos et al. \citep{roos2024social} aims to represent the social dynamics through mediated communication. This model explains that the behavior of users is constrained by three elements: Personal Possibilities, Technical Possibilities, and Social Possibilities, and results in a Situated Social Structure. The authors explain that interpersonal interaction is inherently social and collaborative, and that the smoothness of their interaction is used as evidence to update their beliefs about the relationship with others. Consequently, disruptions caused by technical issues could lead to social expectations being disrupted about the partner’s interaction, which raises questions about common goals and identity as a collective. This model puts the social context into perspective alongside technical and personal ones.

Finally, for social interactions in VR, the feeling of “being in the VE together” (also known as co-presence) in Schroeder study \citep{schroeder2006being}, can be related to social interactions. Indeed, the Cooperative Model of Human Communication of Tomasello \citep{tom2008} can only occur if each individual considers the other as a partner for collaboration. Therefore, if a virtual avatar does not provide the feeling that it is really controlled by a human, it could be challenging to experience collaboration or social interaction. Therefore, Schroeder presents a model to represent the continuum of connected presence, which can be applied to various devices that allow communication. A distinction is made between co-presence and connected presence based on the shared environment that characterizes connected presence. This model considers co-presence as the feeling of being with others through mediated communication devices in which users do not share the same environment (i.e., phone call); and uses the term connected presence when the environment is shared (i.e., social VE and computer). This distinction is made in the case of asymmetric collaboration. In our study concerning social interactions in VR, participants will use a symmetrical mediated device. Therefore, we will use the term "co-presence" for the feeling of being in the VE together, as described by Biocca et al. \citep{biocca2003toward}.

Therefore, the co-presence level, measured with a questionnaire \citep{biocca2003toward}, can be seen as an indicator of the perceived quality of the social interaction. Co-presence represents the feeling that avatars are the representations of their human users, that the avatars \textbf{are} the humans. According to the Theory of Mind \citep{gallagher2003functional} and previous communication models presented above, a qualitative social interaction only occurs when we consider the other as human. In other words, participants in VR perceive social interaction in VR as equivalent in quality to face-to-face interaction. Therefore, to study co-presence in VR, it is necessary to conduct social interactions, which can be seen as a reciprocal influence in a situated context \citep{stebe2007risques}. A meta-analysis found that social influence with avatars or agents only emerges in collaborative or competitive tasks, and does not occur in neutral ones \cite{fox2015avatars}. In both collaboration and competition, the simultaneous orientation of the actors is directed towards the same object of the collaboration or the competition. Although co-presence can be measured in any situation or task, collaboration and competition tasks allow us to control the simultaneous orientation of actors on the same subject and on each other, thereby reproducing Newcomb's model \citep{newcomb1953approach}.


\section{Koinos method: applying communication models in VR-mediated systems}

\begin{figure}
    \centering
    \includegraphics[width=0.7\linewidth]{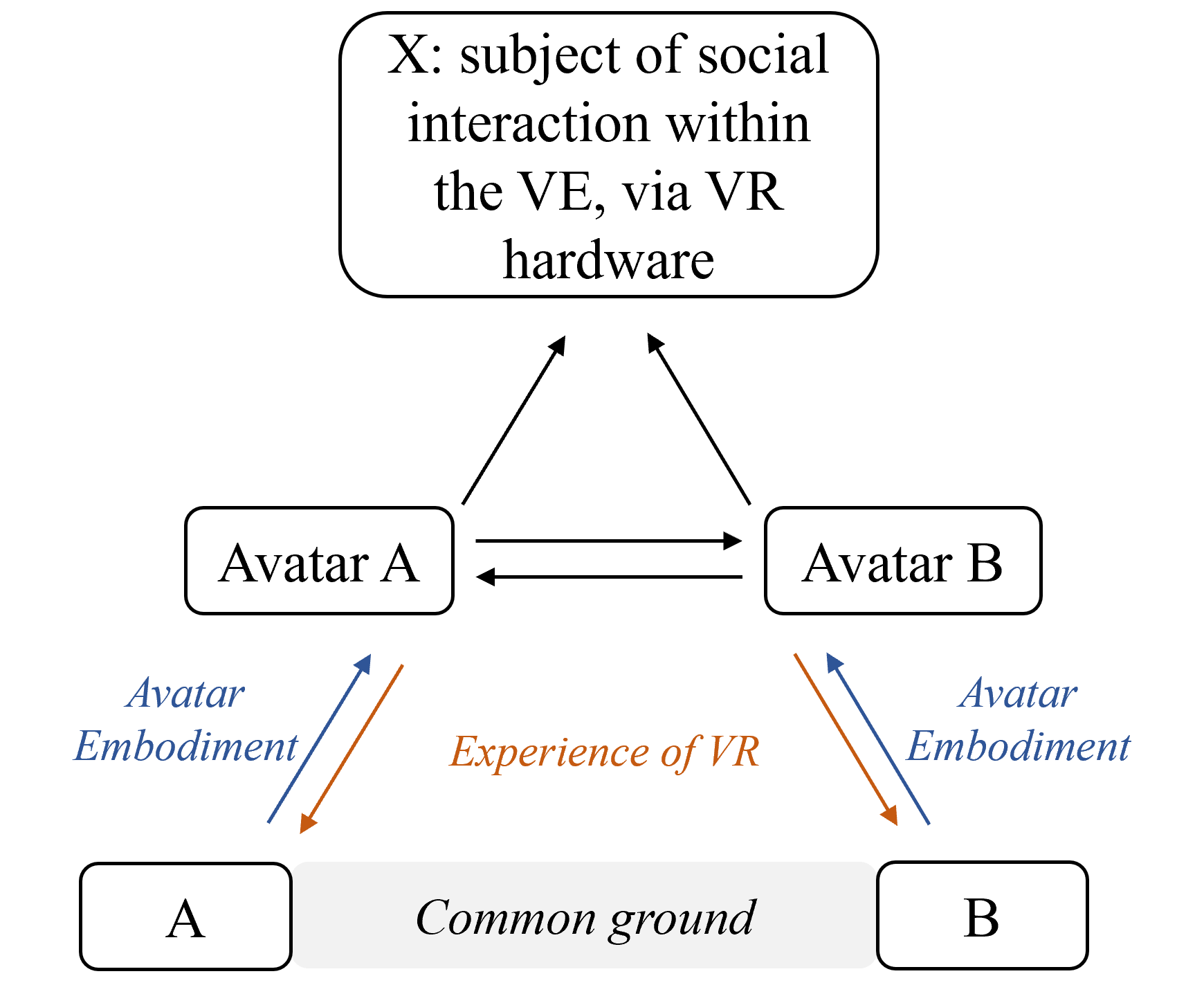}
    \caption{VR social interaction dynamics viewed through communication models: Newcomb’s ABX model \citep{newcomb1953approach} and Cooperative model of Human Communication \citep{tom2008}.}
    \label{fig:comVR}
\end{figure}

To better understand the social interaction dynamics occurring in VR, we suggest applying Newcomb’s ABX model. X will correspond to the message transmitted, which can be related to a task within a VE, through the VR hardware. The relationship between A and B will therefore be mediated; the communication between the interactors will take place through their respective avatars. However, regarding the cooperative model of human communication, a link still exists between A and B: the common ground. Both communication models are combined in a single model depicted in Figure \ref{fig:comVR}, which describes the social interaction dynamics in VR, with a focus on the embodied avatar.

This model enables us to understand that, even if communication is mediated by avatars, a link still exists between users A and B through the common ground. Based on past literature, we can propose that this common ground may have different “strengths” that can influence the communication smoothness and the social interaction.

Resulting from Tomasello’s theory \citep{tom2008}, we propose a grid that allows for estimating the common ground “strength” (Table \ref{Fig1}).

\begin{table}
\resizebox{\columnwidth}{!}{%
\begin{tabular}{ccccccccc}
\multicolumn{9}{l}{\textit{Only one possibility in each of the three columns:}}\\
\multicolumn{3}{c|}{\textbf{\begin{tabular}[c]{@{}c@{}}Is the common ground \\ based on:\end{tabular}}}                                                                                                    & \multicolumn{3}{c|}{\textbf{\begin{tabular}[c]{@{}c@{}}Is the situation where\\ the common ground \\ take place is:\end{tabular}}}                                                               & \multicolumn{3}{c}{\textbf{\begin{tabular}[c]{@{}c@{}}Is the common ground \\ based on … \\ knowledge:\end{tabular}}}                                                                                                  \\
Nothing                & \begin{tabular}[c]{@{}c@{}}Immediate\\ perceptual\\ environment\end{tabular} & \multicolumn{1}{c|}{\begin{tabular}[c]{@{}c@{}}Shared \\ experience \\ from the past\end{tabular}} & Missing                & \begin{tabular}[c]{@{}c@{}}Based on \\ bottom-up \\ process\end{tabular} & \multicolumn{1}{c|}{\begin{tabular}[c]{@{}c@{}}Based on \\ top-down \\ process\end{tabular}} & \begin{tabular}[c]{@{}c@{}}Absent \\ common\end{tabular} & \begin{tabular}[c]{@{}c@{}}Explicit\\ common\\ cultural\\ knowledge\end{tabular} & \begin{tabular}[c]{@{}c@{}}Implicit \\ common \\ knowledge\end{tabular} \\ \hline
\multicolumn{1}{c|}{0} & \multicolumn{1}{c|}{1}                                                       & \multicolumn{1}{c|}{2}                                                                             & \multicolumn{1}{c|}{0} & \multicolumn{1}{c|}{1}                                                   & \multicolumn{1}{c|}{2}                                                                       & \multicolumn{1}{c|}{0}                                    & \multicolumn{1}{c|}{1}                                                           & 2                                                                       \\ \hline
\multicolumn{3}{c|}{a}                                                                                                                                                                                     & \multicolumn{3}{c|}{b}                                                                                                                                                                           & \multicolumn{3}{c}{c}                                                                                                                                                                                                  \\ \hline
\multicolumn{9}{c}{\textbf{Total: a + b + c}} 
\end{tabular}
}
\caption{Proposed Evaluation Grid of the common ground strength between individuals. The lowest value is 0 and the highest is 6.}\label{Fig1}
\end{table}

Tomasello \citep{tom2008} explains that the common ground will be stronger when individuals share experience from the past, when the situation is based on a top-down process, and when they share implicit common knowledge. The values of “1” or “2” were chosen according to these results. For instance, a value of “2” is associated with a shared experience from the past, and a value of “1” is assigned to the immediate perceptual environment. The grid allows for the evaluation of the strength of the common ground in a given situation and individuals, which can change over time. The “0” value is set in the cases where the chosen common ground item does not exist. 

According to the proposed grid, the highest value for the common ground is 6, when the common ground is based on shared experience from the past, when the situation implements a top-down process, and when the common knowledge is based on implicit common knowledge. For instance, co-workers who work together for several years on a common project, and know what the other knows about this project, induce the maximum value of common ground. Conversely, the lowest value can be 0; therefore, no common ground is present. A low value can be identified, for instance if strangers walk in a street, and hear a strange noise, it could induce a value of 3 (one in each three categories), or two if the strangers are tourists from different countries for instance (explicit common cultural knowledge is then rated as 0).

The proposed methodology aims to identify the various factors influencing social interaction dynamics in a VR application (Figure \ref{fig:comVR}). The use of the common ground evaluation grid (Table \ref{Fig1}) allows for the estimation of the strength of the common ground, and a better understanding of the link between participants and the social interaction dynamics that occur. We propose to name this methodology the \textbf{Koinos Method}. The term Koinos (\foreignlanguage{greek}{κοινός}) is derived from ancient Greek, signifying “common”, “shared” or “collective”. In the context of this methodology, Koinos represents the underlying principles of shared understanding, collaborative interaction, and collective experience; elements that are central to the dynamics of social interaction in VR.

\subsection*{Toward a prediction of the co-presence level?}
From past research and our proposed Koinos Method, we question if the social interaction dynamics in VR could be predicted, especially the co-presence level. Indeed, by considering that co-presence reflects the perceived quality of the social interactions, the fact that with a high common ground, individuals do not need complex communication, and that the social interactions happen in a constrained VE, we wonder if identifying and studying each of these factors could help predict the co-presence level. The social interactions allowed and implemented in the VE seems to have an impact in this prediction, thus we will use the term of communication cues to define the communication means that are included in the VR application, that could be verbal and non-verbal, such as voice, text chat, eyes or face tracking and animation. 

Therefore, if users share a strong common ground, do they require complex communication cues in VR to understand each other and achieve high-quality social interactions? To explore this, we analyse two VR experiments, each addressing distinct research questions. However, by utilizing the same collaborative task, these experiments allow for comparative analysis with different virtual partners, either by altering social representation or avatar appearance.

\section{Applying the Koinos Method on collaborative VR experiments}
\subsection{VR experiment 1: Effect of the social representation on co-presence and behavior}
A first study was designed to explore whether it is possible to induce different co-presence levels with the social representation theory, based on Abric \citep{abric1989etude} and Poinsot et al. \citep{poinsot2022effect}, and an online survey on the social representation of AI. We have explored the following research questions: Will co-presence be affected by different social representations of the virtual collaborator? Will participants collaborate in the same way with a \textit{Human} collaborator or with an \textit{AI}-based one? Will user's behavior be correlated to the co-presence level they experience? 

\subsubsection{Experimental design}
The main experimental condition was to induce different levels of co-presence, in order to study what other factors could be related to it. Participants experienced two VR sessions, one with a \textit{Human}-controlled avatar and the other with an \textit{AI}-controlled avatar, in random order. In reality, in both cases, participants completed the experiment together without knowing it, being the \textit{Human} or \textit{AI} partner of each other without knowing it. They were separated in different rooms without the possibility of seeing, talking, or hearing each other. 

\begin{figure}
    \centering
    \includegraphics[width=0.75\linewidth]{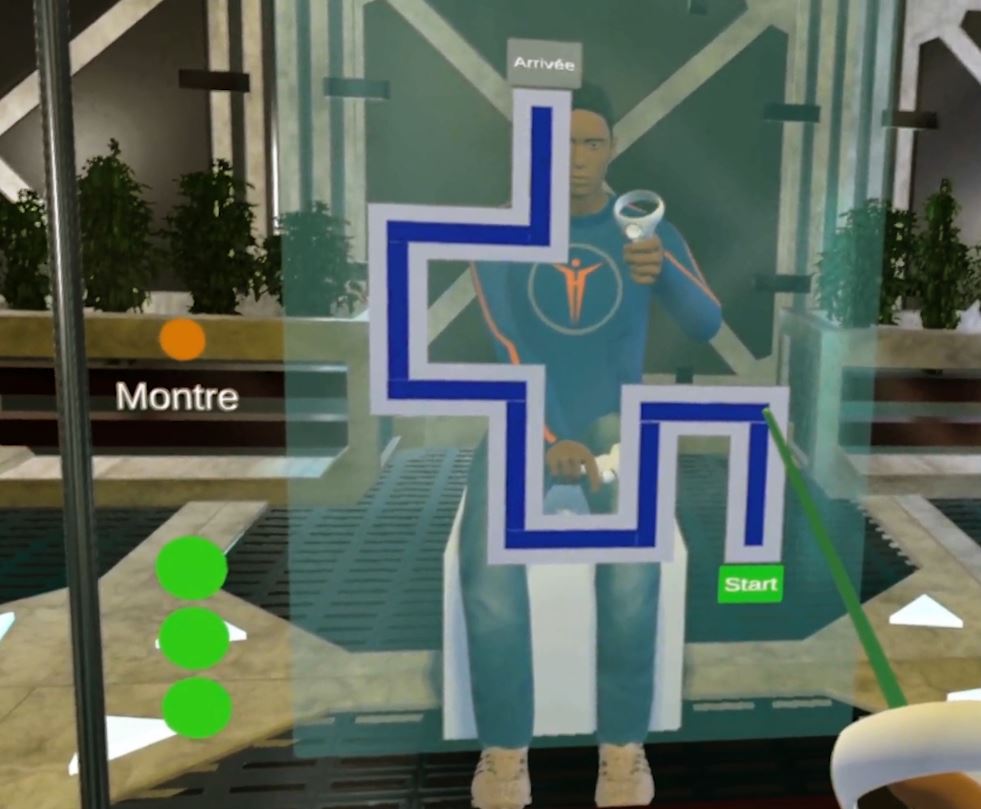}
    \includegraphics[width=0.75\linewidth]{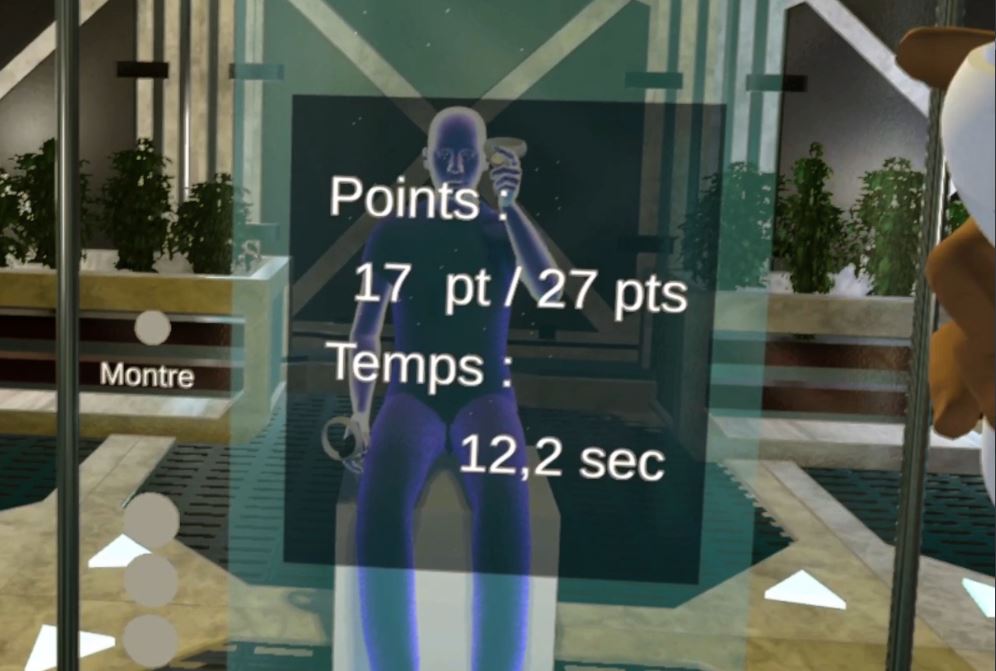}
    \caption{Screenshots of the task in VR, completed with a Human virtual partner on the left, with an AI virtual partner on the right.}
    \label{fig:xp1}
\end{figure}

\begin{figure}
    \centering
    \includegraphics[width=1\linewidth]{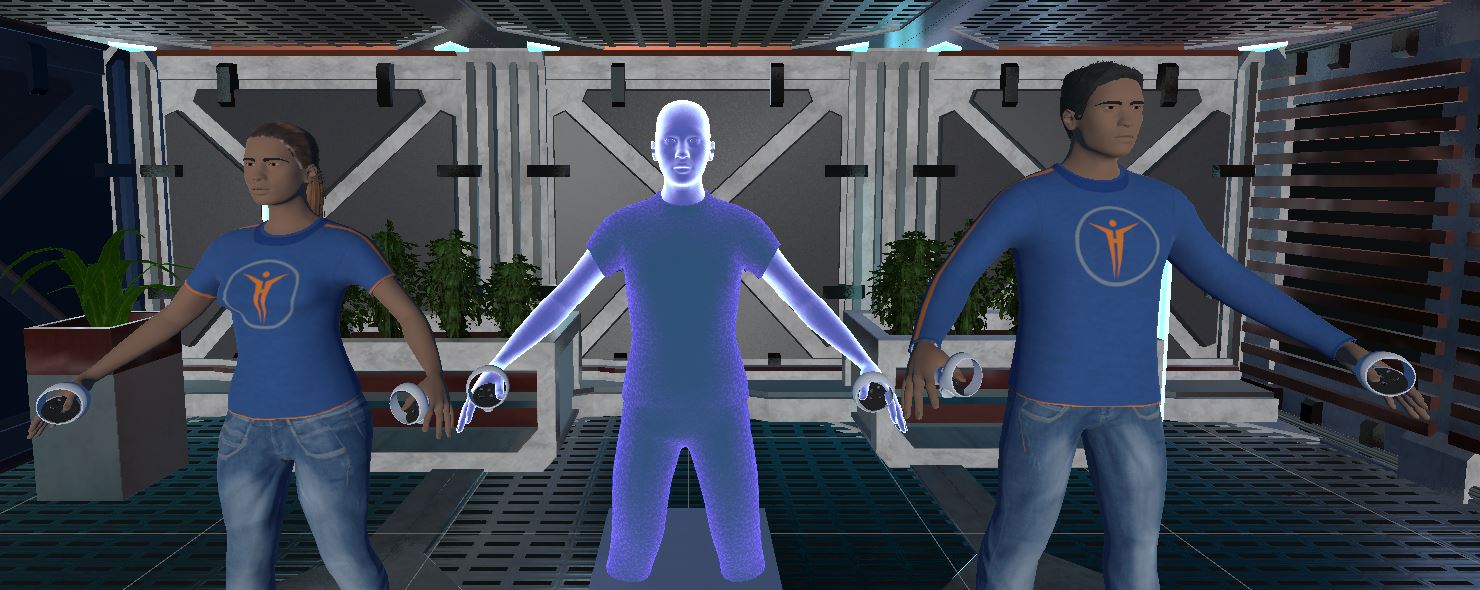}
    \caption{Three different representations used in the first experiment for the virtual collaborator: from left to right, Female human-like, AI, Male human-like.}
    \label{fig:xp1_avatars}
\end{figure}

A collaborative task was developed in VR for Oculus Quest 2, in which the goal was to gain points by completing patterns in collaboration, taking into account the speed and accuracy. Each participant had a specific task: either following a pattern with the hand or reproducing the movement of the collaborator (Figure \ref{fig:xp1}). In order to generate social representations of each entity, the avatar appearance was congruent to the entity avatar represented, i.e., a human shape for the \textit{Human}-controlled avatar (Figure \ref{fig:xp1_avatars}, left and right), and an avatar representing an AI for the \textit{AI}-controlled avatar (Figure \ref{fig:xp1_avatars}, center). Facial expressions were not included and participants could communicate with gestures between two completions of the pattern.

Previous to the experiment, an online survey was conducted in order to identify which visual representation is associated with AI, and what words or sentences are associated with AI. A total of 105 answers were analyzed, allowing us to identify that the visual representation of AI is a mix of the algorithm idea with a human representation, such as a mathematical formula with a humanoid robot that looks like it is thinking. The social representation of AI results in terms such as “algorithms”, “help”, or “non-human” terms.

\subsubsection{Procedure and participants}
The experiment was presented to both participants together, explaining that they would have to compare their impressions of the \textit{Human}-controlled avatar and \textit{AI}-controlled avatar conditions. 
In the VR application, participants began by filling in a 2D grid assessing their emotion of Russel et al. \citep{russell1989affect}, and completed a training of the task. The experimenter connected participants together, and they conducted the task. They ended the VR session by filling out a questionnaire that contained: a Fast Motion Sickness question of Keshavarz \& Hecht \citep{keshavarz2011validating} and the social presence questionnaire of Harms \& Biocca \citep{harms2004internal} on a touch screen tablet.
This study has recorded 52 participants from various sectors and jobs roles. The average age was $M=32.44, SD = 16.12$ and included 14 women and 38 men. This study was approved by an ethics committee from Bourgogne Franche-Comté University, n°CERBFC-2023-05-25-028.

\subsubsection{Results}
\begin{figure}
    \centering
    \includegraphics[width=0.75\linewidth]{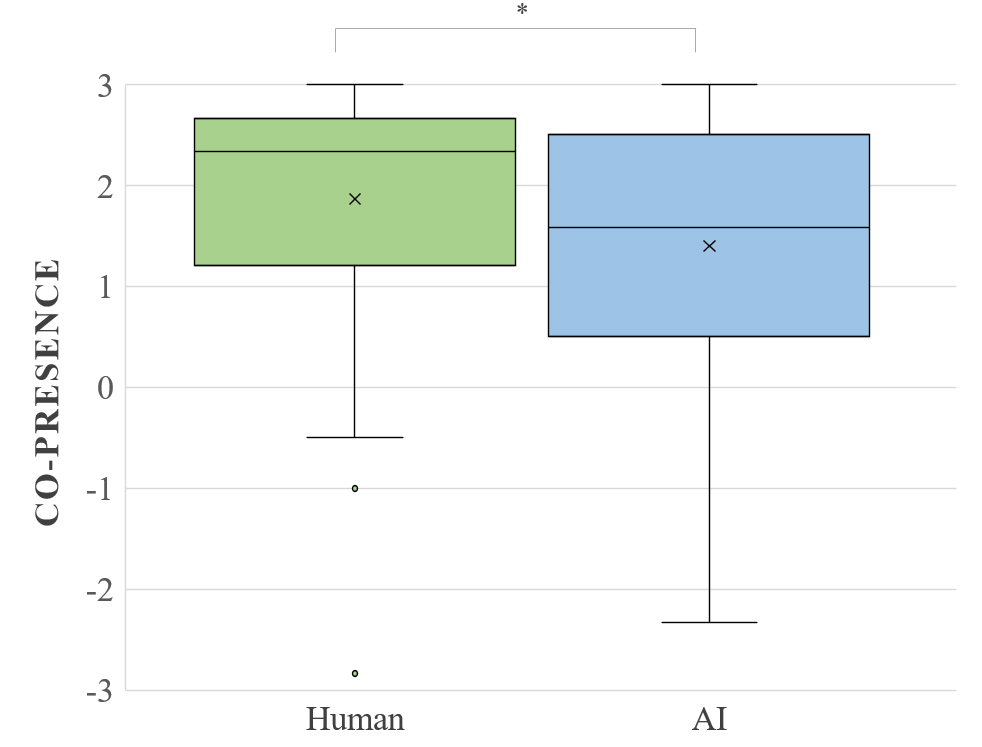}
    \caption{Co-presence results between \textit{Human} and \textit{AI} avatars experimental conditions with 52 participants divided equally between the two conditions.}
    \label{fig:xp1Copres}
\end{figure}

The data were first tested for normality and homoscedasticity, which revealed that the use of non-parametric tests was relevant.
The first main result of this study is the difference in co-presence level between the \textit{Human} ($M = 1.87, SD = 1.21$) and \textit{AI} avatars experimental conditions ($M = 1.41, SD = 1.26$, with $Z = 2.21, p = .027$) from a Wilcoxon signed-rank test (Figure \ref{fig:xp1Copres}).

The second result is the absence of differences in the task performance, either in score and time completion, between the \textit{AI} ($M=191.34, SD=50.14$) and \textit{Human} ($M=191.96, SD=50.04$) avatars conditions, $Z=.30, p=.76$.
Finally, Spearman correlation tests indicate that the three variables showing the highest Correlation Factor ($CF$) with co-presence are Perceived Behavioral Interdependence ($CF=.545, p<.001$) (Harms \& Biocca, \citep{harms2004internal}), Looking at the Avatar's Body ($CF=.486, p<.001$), and Perceived Message Understanding ($CF=.442, p<.001$) (Harms \& Biocca, \citep{harms2004internal}).

\subsubsection{Discussion}
The study was analyzed regarding the AI social representation of the online survey, and it shows that co-presence and behavior were influenced by it. The co-presence level is significantly lower in the \textit{AI}-controlled avatar condition than with the \textit{Human}-controlled avatar.
The social representation of AI seems relevant to explain this result. An AI is seen as a help, as an assistant, participants may have expected an efficient behavior to complete the task, and saw it as a non-human entity. Therefore, the feeling of being with someone in VR with an \textit{AI}-controlled avatar is significantly lower than with a \textit{Human}-controlled avatar, even told that this human is not known.


In addition, correlation tests did reveal three factors that can be related to communication (e.g., Perceived Behavior Interdependence, Looking at the avatar’s body, Perceived Message Understanding).

\subsection{VR experiment 2: Does the avatar appearance really matter?}
Given the results of the first experiment, in which we varied the social representations of the virtual collaborators by using a congruent avatar appearance, we explored another aspect of social interactions: the impact of the avatar appearance on social interactions. Using the same task as in the first experiment, the main condition involved changing the avatar appearance with the same virtual partner; participants completed the task together, knowing that they were collaborating, in contrast to the first experiment.

\subsubsection{Experimental design}
\begin{figure}
    \centering
    \includegraphics[width=1\linewidth]{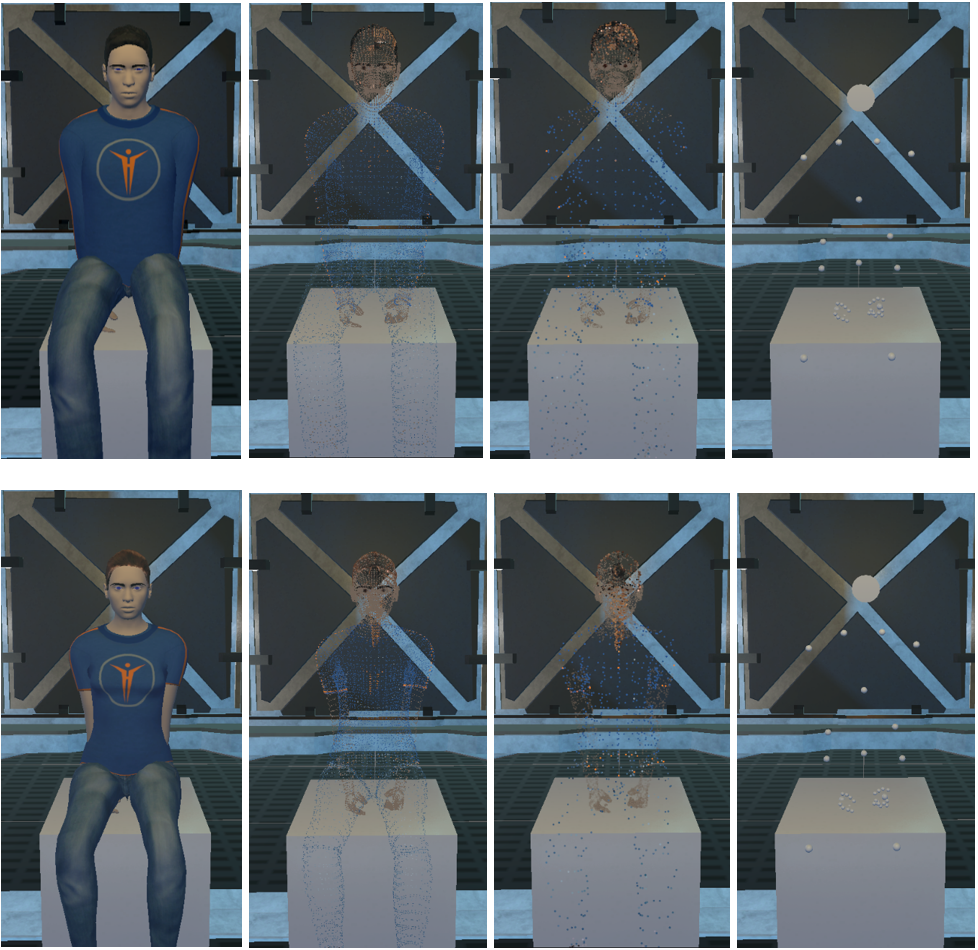}
    \caption{The four avatars type developed and used in the experiment for male (top) and female (down). From left to right: Meshed avatar (M), the High number of point avatar of 70 000 points (HP), the Medium number of point avatar of 10 000 points (MP) and the Low number of point avatar of 60 points (LP).}
    \label{fig:xp2Avatars}
\end{figure}

A similar experimental design was used as the one for the previous VR experiment. Four avatar appearances were compared: one meshed avatar, the most used according to Weidner et al. \citep{weidner2023systematic}, and three point-cloud avatars. Point-cloud avatars were used in the objective to induce lower social bias associated with the avatar's appearance, based on (Hudson \& Hurter \citep{hudson2016avatar} and Yee \& Bailenson \citep{yee2007proteus}, while allowing an anthropomorphic representation and reproduction of human behavior. Due to organizational constraints, participants tested two avatars, divided into two groups. One group tested in random order, the meshed (M, Figure \ref{fig:xp2Avatars} up-left) avatar and the avatar with the highest number of point avatar (HP, 70 000 points, Figure \ref{fig:xp2Avatars} up-right), allowing to recognize the shape of the meshed avatar. The second group had tested in random order, the medium number of points (M, 10 000 points, Figure \ref{fig:xp2Avatars} down-left) and the lowest number of points (LP, 60 points, Figure \ref{fig:xp2Avatars} down-right) avatars. Before the task began, participants underwent an embodiment phase in front of a virtual mirror, following five instructions of 40 seconds each.

\subsubsection{Procedure and participants}
A similar procedure was conducted as in experiment 1. However, some differences can be noted: participants were told to collaborate together in both VR sessions, an embodiment phase was added in the VR application, as well as an embodiment questionnaire (Gonzalez-Franco \& Peck, \citep{gonzalez2018avatar}).
This study has recorded 48 participants, mainly students from business school, engineering school, and computer science programs. The average age was $M=21.17, SD = 3.34$ and included 16 women and 32 men. This study was approved by an ethics committee from Bourgogne Franche-Comté University, n°CERBFC-2024-06-03-022.

\subsubsection{Results}
\begin{figure}
    \centering
    \includegraphics[width=0.75\linewidth]{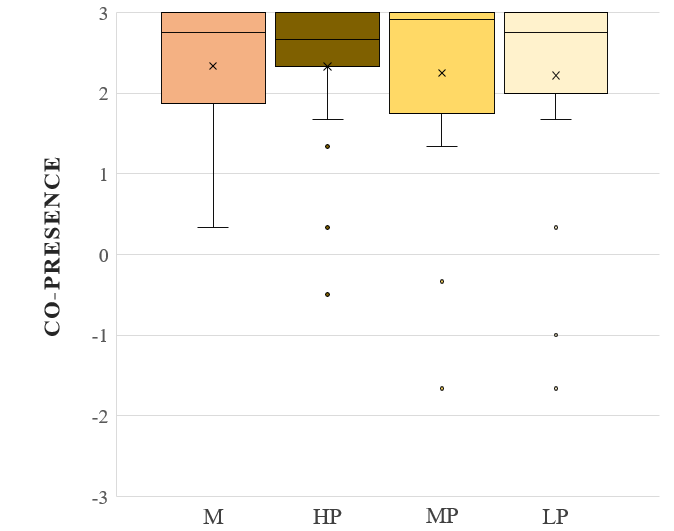}
    \caption{Co-presence results between the four avatars experimental conditions with 48 participants divided equally between the conditions.}
    \label{fig:xp2Copres}
\end{figure}

The data were first tested for normality and homoscedasticity, which revealed that the use of non-parametric tests was relevant.
First, a high co-presence level was noted for each avatar, with a ceiling effect (Figure \ref{fig:xp2Copres}). The average values and Standard deviation of each avatar are: $M=2.33, SD=.94$ for the M avatar, $M=2.33, SD=.90$ for the HP avatar, $M=2.25, SD=1.18$ for the MP avatar and $M=2.22, SD=1.27$ for the LP avatar. A mixed linear model analysis did not show any effect of the avatar condition on the co-presence level ($F=.03, p=.9$7).

Secondly, the results did not represent significant difference in task performance regarding the avatar condition, either during the training ($F=.50, p=.68$) or during the task ($F=.30, p=.82$).

A Spearman correlation test did not reveal any significant correlation the between embodiment sub-categories (e.g., Body Ownership, Agency and Motor Control, and External appearance) results and the co-presence level. However, the observed ceiling effect may have impacted these results. It is interesting to note that a significant effect is present between the four avatars regarding the embodiment sub-categories in “Agency and Motor control” ($F=3.68, p=.02$) and “External appearance” ($F=7.60, p <.001$).

Finally, the Spearman test indicates that the three variables showing the highest Correlation Factor ($CF$) with co-presence are Perceived Behavioral Interdependence ($CF=.641, p<.001$), Emotional valence before VR ($CF=.398, p <.001$) and Perceived Message Understanding ($CF=.347, p<.005$).

\subsubsection{Discussion}
This study was analyzed with communication models, using the ABX model of Newcomb \citep{newcomb1953approach} and the Cooperative Model of Human Communication of Tomasello \citep{tom2008}. The participants having a strong common ground, the communication cues were limited, and the co-presence level was high for each avatar. A ceiling effect was observed on the co-presence level results. Therefore, the avatar's appearance did not impact the social interaction, and thus had less impact than the common ground. This study compared abstract avatars, with a point cloud representation, with human body representation, using the same tracking system and avatar animation (inverse kinematic). This analysis suggests investigating the impact of communication cues on the social interaction perceived quality, as well as investigating abstract avatars that could reproduce human behavior with lower social bias than meshed avatars. 

\subsection{Implementing the Koinos Method on the VR experiments}
The application of the communication models in the second experiment allowed us to better understand the social interaction dynamics in this collaborative task. Also, it helped us propose that the avatar's appearance may have less impact on the social interaction than a strong common ground.

We propose applying the same analysis to the first VR experiment, which utilizes the same VR application. The participants communicated through their avatars and the VR hardware, Virtual Environment, and User Interface. However, contrary to the second experiment, participants did not know with whom they completed the task. The experimenter explained that they would complete the task either with a human, a student in another lab that they did not know, or with an avatar controlled by an AI. In reality, participants completed the task together. The results indicated lower co-presence levels in the \textit{AI}-controlled avatar condition than in the \textit{Human}-controlled avatar condition. We first apply the common ground evaluation grid for the Human condition (Table \ref{tab2}):

\begin{table}
\resizebox{\columnwidth}{!}{%
\begin{tabular}{ccccccccc}
\multicolumn{9}{l}{\textit{Only one possibility in each of the three columns:}}\\
\multicolumn{3}{c|}{\textbf{\begin{tabular}[c]{@{}c@{}}Is the common ground \\ based on:\end{tabular}}}                                                                                                    & \multicolumn{3}{c|}{\textbf{\begin{tabular}[c]{@{}c@{}}Is the situation where\\ the common ground \\ take place is:\end{tabular}}}                                                               & \multicolumn{3}{c}{\textbf{\begin{tabular}[c]{@{}c@{}}Is the common ground \\ based on … \\ knowledge:\end{tabular}}}                                                                                                  \\
\sout{Nothing}                & \begin{tabular}[c]{@{}c@{}}Immediate\\ perceptual\\ environment\end{tabular} & \multicolumn{1}{c|}{\begin{tabular}[c]{@{}c@{}}\sout{Shared} \\ \sout{experience} \\ \sout{from the past}\end{tabular}} & \sout{Missing}                & \begin{tabular}[c]{@{}c@{}}\sout{Based on} \\ \sout{bottom-up} \\ \sout{process}\end{tabular} & \multicolumn{1}{c|}{\begin{tabular}[c]{@{}c@{}}Based on \\ top-down \\ process\end{tabular}} & \begin{tabular}[c]{@{}c@{}}\sout{Absent} \\ \sout{common}\end{tabular} & \begin{tabular}[c]{@{}c@{}}Explicit\\ common\\ cultural\\ knowledge\end{tabular} & \begin{tabular}[c]{@{}c@{}}\sout{Implicit} \\ \sout{common} \\ \sout{knowledge}\end{tabular} \\ \hline
\multicolumn{1}{c|}{\sout{0}} & \multicolumn{1}{c|}{1}                                                       & \multicolumn{1}{c|}{\sout{2}}                                                                             & \multicolumn{1}{c|}{\sout{0}} & \multicolumn{1}{c|}{\sout{1}}                                                   & \multicolumn{1}{c|}{2}                                                                       & \multicolumn{1}{c|}{\sout{0}}                                    & \multicolumn{1}{c|}{1}                                                           & \sout{2}                                                                       \\ \hline
\multicolumn{3}{c|}{a = 1}                                                                                                                                                                                     & \multicolumn{3}{c|}{b = 2}                                                                                                                                                                           & \multicolumn{3}{c}{c = 1}                                                                                                                                                                                                  \\ \hline
\multicolumn{9}{c}{\textbf{Total: 1 + 2 + 1 = 4}} 
\end{tabular}
}
\caption{Common ground evaluation grid apply on the first VR experiment, in the Human-controlled avatar condition.}\label{tab2}
\end{table}

With the human partner, due to the collaborative task, we propose that common ground is based on an immediate perceptual environment, on a top-down process, and on explicit common cultural knowledge, the human partner being presented as a student in another laboratory of the same country. 

With the AI partner, the use of the grid may not be relevant. Indeed, the AI entity is a non-human entity. Therefore, AI is seen as an entity with which it is possible to cooperate to complete a task (related to the social presence presented in the first experiment). However, we cannot argue whether this entity is perceived as equal to a human in terms of cooperation. Indeed, with a human, the Theory of Minds of Gallagher \& Frith \citep{gallagher2003functional} explains that human beings are considered as entities having their own thoughts, beliefs, values, goals, are therefore relevant as cooperative entities and correspond to the Cooperative Model of Human Communication of Tomasello \citep{tom2008}. With the actual knowledge and literature, we may say that the common ground is null with a non-human entity. However, this assumption requires further study, considering that humans are increasingly accustomed to interacting with non-human entities, such as AI, and that their behavior is becoming more similar to human behavior. 

Regarding the results of the AI social representations online survey and the resulting data of the experiment, social representations of the AI entity impact the interaction. However, social representations are indeed linked to common ground: for instance, in a given situation, even if the common ground can be strong, the social representations of the other person or entity influence behaviors. In other words, the common ground can be considered null with the AI entity and higher with the human entity, whereas the communication cues are the same in both cases and the same in the second experiment, therefore limited to only animation of the hands and the head movements.

From the two experiments, the co-presence level is higher in the second experiment than in the first human condition. Despite differential elements in the experimental procedures, the common ground is stronger in the second experiment than in the first; participants knew with whom they would complete the task, while they thought that they did not know their partner in the first experiment. We apply the common ground evaluation grid on the second experiment (Table \ref{tab3}):

\begin{table}
\resizebox{\columnwidth}{!}{%
\begin{tabular}{ccccccccc}
\multicolumn{9}{l}{\textit{Only one possibility in each of the three columns:}}\\
\multicolumn{3}{c|}{\textbf{\begin{tabular}[c]{@{}c@{}}Is the common ground \\ based on:\end{tabular}}}                                                                                                    & \multicolumn{3}{c|}{\textbf{\begin{tabular}[c]{@{}c@{}}Is the situation where\\ the common ground \\ take place is:\end{tabular}}}                                                               & \multicolumn{3}{c}{\textbf{\begin{tabular}[c]{@{}c@{}}Is the common ground \\ based on … \\ knowledge:\end{tabular}}}                                                                                                  \\
\sout{Nothing}                & \begin{tabular}[c]{@{}c@{}}\sout{Immediate} \\ \sout{perceptual} \\ \sout{environment}\end{tabular} & \multicolumn{1}{c|}{\begin{tabular}[c]{@{}c@{}}Shared \\experience \\from the past\end{tabular}} & \sout{Missing}                & \begin{tabular}[c]{@{}c@{}}\sout{Based on} \\ \sout{bottom-up} \\ \sout{process}\end{tabular} & \multicolumn{1}{c|}{\begin{tabular}[c]{@{}c@{}}Based on \\ top-down \\ process\end{tabular}} & \begin{tabular}[c]{@{}c@{}}\sout{Absent} \\ \sout{common}\end{tabular} & \begin{tabular}[c]{@{}c@{}}\sout{Explicit} \\ \sout{common} \\ \sout{cultural} \\ \sout{knowledge}\end{tabular} & \begin{tabular}[c]{@{}c@{}}Implicit \\ common \\ knowledge\end{tabular} \\ \hline
\multicolumn{1}{c|}{\sout{0}} & \multicolumn{1}{c|}{\sout{1}}                                                       & \multicolumn{1}{c|}{2}                                                                             & \multicolumn{1}{c|}{\sout{0}} & \multicolumn{1}{c|}{\sout{1}}                                                   & \multicolumn{1}{c|}{2}                                                                       & \multicolumn{1}{c|}{\sout{0}}                                    & \multicolumn{1}{c|}{\sout{1}}                                                           & 2                                                                       \\ \hline
\multicolumn{3}{c|}{a = 2}                                                                                                                                                                                     & \multicolumn{3}{c|}{b = 2}                                                                                                                                                                           & \multicolumn{3}{c}{c = 2}                                                                                                                                                                                                  \\ \hline
\multicolumn{9}{c}{\textbf{Total: 2 + 2 + 2 = 6}} 
\end{tabular}
}
\caption{Common ground evaluation grid apply on the second VR experiment; ; the virtual partner is known.}\label{tab3}
\end{table}

A score of 6 results from the use of the common ground evaluation grid for the second experiment, while a score of 4 for the first experiment in the \textit{Human}-controlled avatar condition, and a (supposed) score of 0 for the \textit{AI}-controlled avatar condition. The co-presence level is then correlated to these scores, despite the use of the same limited communication cues. Therefore, with a high common ground between participants, it is not necessary to have complex communication cues for participants to experience a high co-presence level. On the other hand, having a lower common ground is associated with a lower co-presence level, and even lower with a non-human partner.

The results also showed a significant positive correlation between co-presence and Perceived Behavioral Interdependence (PBI) and Perceived Message Understanding (PMU) in both experiments. 

In other words, when participants perceived the interdependence of behavior, they also experienced a high level of co-presence, and this was similar for PMU. Based on communication theories, behavioral interdependence and message understanding are essential to experiencing smooth social interaction. Therefore, this task proposed relevant communication cues (exclusively non-verbal) regarding the objectives of the task, and it engaged participants together. 

\section{General Discussion: Toward a Co-presence equation?}
The application of the proposed Koinos Method allows us to identify a pattern that seems to link co-presence, common ground and communication cues, and are summarized in the Table \ref{tab4}. 

\begin{table}
\resizebox{\columnwidth}{!}{%
\begin{tabular}{|l|l|c|l|}
\hline
\textbf{VR   experiment}                                                     & \textbf{Co-presence   level}                                                                  & \multicolumn{1}{l|}{\textbf{\begin{tabular}[c]{@{}l@{}}Common   \\ ground value\end{tabular}}} & \textbf{\begin{tabular}[c]{@{}l@{}}Communication   \\ cues\end{tabular}} \\ \hline
\begin{tabular}[c]{@{}l@{}}N°1 –   Human \\ (unknown) condition\end{tabular} & \textit{\begin{tabular}[c]{@{}l@{}}High \\    \\ (M   = 1.87, SD = 1.21)\end{tabular}}        & 4                                                                                              & Limited                                                                  \\ \hline
N°1 –   AI condition                                                         & \textit{\begin{tabular}[c]{@{}l@{}}Medium   \\    \\ (M   = 1.41, SD = 1.26)\end{tabular}}    & 0                                                                                              & Limited                                                                  \\ \hline
\begin{tabular}[c]{@{}l@{}}N°2 – Known   \\ human\end{tabular}               & \textit{\begin{tabular}[c]{@{}l@{}}Very   high \\    \\ (M =   2.22, SD = 1.27)\end{tabular}} & 6                                                                                              & Limited                                                                  \\ \hline
\end{tabular}
}
\caption{Summary of the co-presence levels regarding the common ground evaluation and communication cues of the two VR experiments} \label{tab4}
\end{table}

From these results and the literature review, we propose this equation:

\begin{equation}
Co-Presence = Common \ ground \ + \ Communication \ cues
\end{equation}

The two experiments allowed us to verify the variation of common ground; however, the communication cues were the same. The results of the second experiment suggest that one of the experimental conditions used by Gamelin \citep{gamelin2021point} and Yu \citep{yu2021avatars} has no effect (appearance), implying that the observed differences in their studies may be attributed to the second condition (the tracking system). Our second experiment varied the avatar appearance (meshed and point-cloud avatars) with the same tracking system, and a high co-presence value resulted for each avatar. Although an experiment is needed to validate this theory, we can suggest that the tracking system has caused the difference in co-presence in both studies of Gamelin \citep{gamelin2021point} and Yu \citep{yu2021avatars}. If it is confirmed, our proposed equation of co-presence can be validated: with the same common ground, changing the communication cues can influence the co-presence level. Regarding non-verbal communication, a more precise, detailed, human-like tracking and animation could increase the co-presence value, and therefore the social interaction perceived experience. Studies about co-presence with agents point in this direction (Sterna et al. \citep{sterna2024behavioral} and Strojny et al. \citep{strojny2020moderators}).

The analyses lead us to the main result of this study: with the same communication cues, a stronger common ground results in a higher co-presence level. Therefore, although this theory needs to be validated, it is possible to use the co-presence equation to increase the social interaction perceived, depending on the task, the relation, the nature of the entity of interactors, and the communication cues. It is possible to use the common ground evaluation grid to evaluate it at first sight, and adapt the necessary communication cues, along with the interactions needed to complete the task or the objective of the VR application.

\subsection{Limitations and Future works}
The establishment of the Co-Presence equation is mainly based on the already known impact of communication cues on co-presence, and on the two VR experiments presented in this paper; therefore, to validate it, a deeper analysis has to be done on different task types and different social interaction situations. Further studies could be done on the common ground evaluation grid, as well as describe communication cues elements.

From our analyses of the VR experiments and the literature, future work could be on three different aspects: exploring different types of abstract avatars, studying the impact of the tracking system on social interaction, and performing advanced studies of social representation of agents.

\subsection{Usage for developers}
This study aimed to use communication models to better understand social interaction in VR, especially regarding the co-presence level, which reflects the perceived quality of the social interaction.

The result of the analysis of the VR experiment led us to propose an equation that can allow us to predict the co-presence level and identify what elements can be changed in the VR application to enable higher co-presence levels. We can propose to developers to use the common ground evaluation grid first to evaluate how strong the common ground between users can be, and in a second step, evaluate if complex communication is necessary (if the common ground is low). Finally, the design of the avatar and the VR application has to be relevant to both complete the task or the objective, and allow communication with others. The second study indicates that the avatar's appearance is not primary on the common ground, if the latter is strong.

\section{Conclusion}
The main result of this study indicates that using communication models to better understand co-presence in VR seems relevant. We propose a methodology that adapts communication models to the specificities of VR systems, with the aim of analyzing the dynamics of social interactions occurring in VR: the Koinos Method. The analysis of the two VR experiments using the Koinos Method indicates that the avatar's appearance does not affect the co-presence level and task completion in this particular task. In addition, our results show that the prejudices and ideas about the virtual collaborator guide the behavior and the perceived experience of social interaction.

Secondly, the analysis of the VR experiments with communication models suggests that the co-presence level depends on the common ground between participants and the communication cues. Co-presence being considered as an indication of the social interaction perceived quality, our results show that with a strong common ground, complex communication cues are not necessary. Conversely, with a light or null common ground, with a non-human partner such as an agent, communication cues have to be more complex in order to experience higher co-presence. In this case, studying the social representation associated with the non-human partner could help identify which factors could bring this non-human entity more "human". In the case of a social interaction with a human entity, an evaluation of the strength of the common ground seems relevant to determine if complex communication cues have to be included, with the objective of bringing the perceived quality of social interactions closer to the FtF interactions. In this case, we proposed a co-presence equation along with the Common Ground Evaluation Grid, allowing us to predict the co-presence level and to identify which factors could help upgrade this level and the social interaction quality.

These results highlight the impact of the relationship between participants on co-presence. Previous studies on co-presence usually vary technological elements without specifying the relationship between participants, especially regarding the common ground shared. Therefore, we suggest that researchers consider using the Common Ground Evaluation Grid presented in this study to assess the level of common ground between participants in future studies.

\section*{CRediT authorship contribution statement}
\textbf{Eloïse Minder:} Conceptualization, Methodology, Formal analysis, Investigation, Writing - original draft, Writing - review \& editing. \textbf{Sylvain Fleury:} Writing - review \& editing, Validation, Supervision. \textbf{Solène Neyret:} Supervision.  \textbf{Jean-Rémy Chardonnet:} Writing - review \& editing, Validation, Supervision.

\section*{Disclosure statement}
No potential conflict of interest was reported by the authors.

\section*{Funding Sources}
This work was supported by the French region of Bourgogne-Franche-Comté under the researcher-entrepreneur program (ICE) grant 2022 Y\_23815.


  \bibliographystyle{elsarticle-num-names} 
  \bibliography{MyBilbio2}






\end{document}